\documentclass[runningheads]{llncs}
\usepackage[T1]{fontenc}
\usepackage{graphicx}
\usepackage{graphicx}
\usepackage{subcaption}
\usepackage{pgfplots}
\pgfplotsset{compat=newest}

\usepackage{graphicx}
\usepackage{cite}
\usepackage{graphicx}
\usepackage{textcomp}
\usepackage{tikz}
\usepackage{bm}
\makeatletter
\AtBeginDocument{\DeclareMathVersion{bold}
\SetSymbolFont{operators}{bold}{T1}{times}{b}{n}
\SetMathAlphabet{\mathrm}{bold}{T1}{times}{b}{n}
\SetMathAlphabet{\mathit}{bold}{T1}{times}{b}{it}
\SetMathAlphabet{\mathbf}{bold}{T1}{times}{b}{n}
\SetMathAlphabet{\mathtt}{bold}{OT1}{pcr}{b}{n}
\SetSymbolFont{symbols}{bold}{OMS}{cmsy}{b}{n}
\renewcommand\boldmath{\@nomath\boldmath\mathversion{bold}}}
\makeatother

\def\BibTeX{{\rm B\kern-.05em{\sc i\kern-.025em b}\kern-.08em
    T\kern-.1667em\lower.7ex\hbox{E}\kern-.125emX}}
\usepackage{fancyhdr}
\usepackage[T1]{fontenc}
\usepackage{fancyhdr}
\usepackage[T1]{fontenc}
\usepackage[linesnumbered,ruled,vlined]{algorithm2e}
\usepackage[utf8]{inputenc}
\usepackage[utf8]{inputenc} 
\usepackage{graphicx} 
\usepackage{pgfplots}
\usepackage{ulem}
\usepackage{multirow}
\usepackage{graphicx}
\usepackage{textcomp}
\usepackage{academicons}
\usepackage{hyperref}
\usepackage{siunitx}
\RequirePackage{csquotes}
\usepackage{booktabs}
\usepackage{makecell}
\usepackage{multirow}
\usepackage{tikz}
\usetikzlibrary{tikzmark}
\usetikzlibrary{patterns}
\usetikzlibrary{calc}
\usepackage{comment}

\usepackage{orcidlink}
\usepackage[hyperfirst=false]{glossaries}
\usepackage{tablefootnote}

\usepackage{graphicx}
\usepackage{bm} 
\usepackage{tikz}
\usepackage{algpseudocode}
\usepackage{pgfplots}
\useunder{\uline}{\ul}{} 
\usepackage{tikz}
\usetikzlibrary{shapes, arrows.meta, positioning}
 \usepackage{algpseudocode}
\usepackage{tikz}
\usepackage{graphicx}
\usepackage{textcomp}
\usepackage{academicons}
\usepackage{hyperref}
\usepackage{siunitx}
\RequirePackage{csquotes}
\usepackage{makecell}
\usepackage{multirow}
\usepackage{tikz}
\usetikzlibrary{tikzmark}
\usetikzlibrary{calc}
\usepackage{comment}
\usepackage{makecell}

\usepackage{orcidlink}
\usepackage[hyperfirst=false]{glossaries-prefix}
\glsdisablehyper
\usepackage[flushleft]{threeparttable}
\usepackage{dblfloatfix}

\usepackage{tabularray}
\UseTblrLibrary{booktabs}

\DeclareSIUnit\flop{\textsc{Fl}\textsc{Op}}
\DeclareSIUnit\cycle{\textsc{Cycle}}
\DeclareSIUnit[per-mode=symbol]\floppersec{\flop\per\second}
\DeclareSIUnit[per-mode=symbol]\flopperjoule{\flop\per\joule}
\DeclareSIQualifier{\doubleprecision}{FP64}
\DeclareSIQualifier{\singleprecision}{FP32}
\DeclareSIQualifier{\halfprecision}{FP16}
\DeclareSIQualifier{\theoretical}{th}
\DeclareSIUnit\pixel{px}
\newcommand{\doubleblind}[1]{} 
\newcommand{\netx}{NetworkX}
\begin{document}
\title{Efficient Accelerated Graph Edit Distance Computation on GPU}
%
%
\author{Adel Dabah\inst{1}\orcidlink{0000-0001-9175-469X} \and
Andreas Herten\inst{1}\orcidlink{0000-0002-7150-2505}}
\authorrunning{A.Dabah et al.}
%
\institute{ \textit{J\"{u}lich Supercomputing Center} \textit{Forschungszentrum J\"{u}lich}  J\"{u}lich, Germany\\ \email{ a.dabah@fz-juelich.de} 
\email{a.herten@fz-juelich.de}}
\maketitle              
\begin{abstract}


Graph representation is a powerful abstraction of real-world objects and relations. Computing the Graph Edit Distance (GED) between graphs is critical in domains such as bioinformatics, machine learning, and pattern recognition. GED measures the minimum number of edit operations required to transform one graph into another. However, the high computational complexity of optimal and near-optimal methods limits their applicability to large-scale graphs, making high-performance parallel GED computation essential. 
To address this, we propose FAST-GED, a fast and scalable open-source framework for GED computation on GPUs. FAST-GED overcomes existing limitations by combining high accuracy with fast execution through GPU-friendly algorithmic design and efficient mapping to GPU hardware, minimizing host-device communication. The implementation is optimized and tested across multiple GPU architectures. 
We validate FAST-GED on real and synthetic datasets with diverse graph sizes and densities. It achieves speedups of several orders of magnitude over the Python \netx{} library while reaching optimal solutions in most cases. Moreover, it outperforms state-of-the-art approximate methods in both accuracy and scalability. 
We show that FAST-GED enables broader adoption of GED-based solutions in real-world applications. 
\end{abstract}


\section{Introduction}
\label{sec:introduction}
Graph computations are essential for problems in social networks, cybersecurity, logistics, and machine learning. Large real-world graphs require high computational power and scalability, making High-Performance Computing (HPC) crucial for advancing graph-based applications. 
Introduced by Sanfeliu and Fu in 1983~\cite{sanfeliu1983distance},
Graph Edit Distance (GED) finds the minimum \emph{cost set} of edit operations: substitution, insertion, and deletion—needed to transform one graph into another~\cite{bunke1983inexact}. The cost of each operation can be adapted per application, making GED a flexible and adaptive tool. 

The GED problem is NP-hard, as the number of possible solutions grows exponentially with the number of vertices~\cite{zeng2009comparing}. Existing methods are computationally intensive, especially for large graphs, and can be categorized into exact and approximate algorithms. Exact approaches such as Depth-First Search (DFS) and A-Star guarantee optimality but are impractical for large datasets. Approximate methods, including K-Best~\cite{kbest_DABAH2021,dabah2022efficient}, VF2~\cite{cordella2004sub}, and Beam Search (BS)~\cite{neuhaus2006fast}, improve efficiency at the cost of accuracy. Yet, scalability and accuracy remain major challenges, particularly for large graphs.

To address these limitations, we propose FAST-GED, a novel GPU-accelerated framework for GED computation that surpasses existing methods in both speed and accuracy. Leveraging the parallel processing capabilities of GPUs~\cite{gpus}, FAST-GED overcomes the trade-off in K-Best methods, where scalability and accuracy are constrained by the parameter $K$. Our framework efficiently maintains both by exploiting GPU parallelism for large-scale GED computation.

FAST-GED operates on a search tree representing all possible edit paths transforming a source graph into a target graph. The tree is built via vertex-based branching, with each node decomposing into smaller subproblems. Similar to Breadth-First Search (BFS), our method explores levels of the search tree, retaining only the best $K$ nodes at each level. After transferring graph data to the GPU, FAST-GED executes three main kernels: 
A branching kernel where each GPU block expands one node and computes its partial edit distances (PEDs). 
A ranking phase to select the best $K$ child nodes without full sorting, using a two-step local and global ranking via atomic operations. 
An update phase that refreshes data structures with the best nodes, avoiding host-device communication and ensuring scalability.
This process repeats until the final level, yielding the minimum-cost edit path.

We evaluated FAST-GED against state-of-the-art methods on real and synthetic datasets with varying graph sizes and densities. Compared to the Python \netx{} library~\cite{networkx}, FAST-GED achieves optimal accuracy in over \qty{90}{\percent} of cases with an order-of-magnitude lower execution time. On real-world datasets, it consistently outperforms Beam Search (BS)~\cite{neuhaus2006fast} and DFS-1~\cite{Zeina_abu_2015exact}. Furthermore, FAST-GED achieves a 300× speedup on an NVIDIA A100 GPU compared to a parallel CPU version on a 48-core AMD EPYC baseline, demonstrating the strong potential of GPU acceleration for GED computation.


The rest of this paper is organized as follows. \hyperref[sec:Problem_definition]{Section~\ref*{sec:Problem_definition}} introduces the problem formulation and notations. \hyperref[sec:related-works]{Section~\ref*{sec:related-works}} reviews related work. \hyperref[sec:k-best]{Section~\ref*{sec:k-best}} presents the design and implementation of FAST-GED. \hyperref[sec:Tests_and_experiments]{Section~\ref*{sec:Tests_and_experiments}} discusses experimental results. \hyperref[sec:apps]{Section~\ref*{sec:apps}} showcase two important applications of FAST-GED, and \autoref{sec:conclusion} concludes the paper.

\section{Problem Formulation} 
\label{sec:Problem_definition}

In the following, we first define some basic concepts and then define the GED problem formally.

\subsection{Graph}
A graph is a structure used to model pairwise relations between objects~\cite{graph_theory_book_west2001introduction}. It contains a set of vertices connected by a set of edges. In this paper, we consider simple undirected labelled graphs, i.e., at most one edge between vertices and without loops. 
 Formally, a labeled graph is denoted by \( G = (V, E, \alpha, \beta) \), where:
\begin{itemize}
    \item  \( V = \{v_1, v_2, \ldots, v_n\} \) is a set of \( n \) vertices.
    \item  \( E = \{e_1, e_2, \ldots, e_m\} \) is a set of \( m \) edges (\( E \subseteq V \times V \)).
    \item  \( \alpha \) is a labeling function for the vertices, \( \alpha : V \rightarrow L_V \).
    \item \( \beta \) is a labeling function for the edges, \( \beta : E \rightarrow L_E \).
\end{itemize}
Here, \( L_V \) and \( L_E \) represent the sets of possible labels for vertices and edges, respectively, which can be integers, real-valued vectors, text, or symbolic labels. 

\subsection{Graph edit distance}
Given two graphs \( g_1 = (V_1, E_1, \alpha_1, \beta_1) \) and \( g_2 = (V_2, E_2, \alpha_2, \beta_2) \), the graph edit distance \( d(g_1, g_2) \) measures the dissimilarity between them. It is defined as the minimum cost of the edit operations needed to transform \( g_1 \) into \( g_2 \):

\[
d(g_1, g_2) = \min_{\{o_1, o_2, \ldots, o_k\} \in \gamma(g_1, g_2)} \sum_{i=1}^{k} c(o_i)
\]

where \( \gamma(g_1, g_2) \) denotes the set of all possible ways transforming \( g_1 \) into \( g_2 \) (\emph{edit paths}) , and \( c(o_i) \) is the cost of the edit operation \( o_i \) \cite{riesen2009approximate}.

\subsection{GED Operations}
\label{subsec:Operations_of_GED}
The GED problem involves transforming one graph \( g_1 \) into another graph \( g_2 \) through a series of edit operations. In this work, we consider a vertex-centric approach in which edit operations are performed on vertices, and edges are implied. 
Let us consider a vertex $v_i \in V_1$ from $g_1$ and a vertex $u_j \in V_2$ from $g_2$, we denote the three edit operations:

\begin{enumerate}
	\item \textbf{Substitution}: $v_i \rightarrow u_j$,  $v_i$ is \emph{substituted} by $u_j$.
	
	\item  \textbf{Deletion}:  $v_i \rightarrow \epsilon$, $v_i$ is \emph{deleted} from $g_1$.
	
	\item \textbf{Insertion}: $\epsilon \rightarrow u_j$, $u_j$ is \emph{inserted} into $g_1$.  
\end{enumerate}

\subsubsection{Implied edges operations}
\label{subsec:Implied_edges_operations}
In the vertex-centric approach, edge operations are implied~\cite{riesen2009approximate}. Therefore, implied edge operations are not applied as independent edit steps on intermediate graphs. Instead, whenever a new vertex operation is applied to the current edit path $\lambda$, we check its incident edges with all previously edited vertices in $\lambda$ and compute the corresponding edge costs (substitution, insertion, or deletion). 
Let us consider two vertices  $v, v'$ from graph $g_1$ and two other vertices $u, u'$ from graph $g_2$ and $\lambda = \{..., v \rightarrow u\}$. If we perform the following edit operation $\{v' \rightarrow u'\}$, three cases of implied edges exist:

\begin{enumerate}
	\item If there is an edge $e_1 (v, v')$ in $g_1$ and there is also an edge $e_2 =(u, u')$ in $g_2$, then $e_1$ is \emph{substituted} by $e_2$, denoted $e_1 \rightarrow e_2$.

	\item If there is an edge $e_1 (v, v')$ in $g_1$ and there is no edge between $u$ and $u'$ in $g_2$, then $e_1$ is \emph{deleted} from $g_1$,  denoted $e_1 \rightarrow \epsilon$.

	\item If there is no edge between $v$ and $v'$ in $g_1$ and there is an edge $e_2 (u, u')$ between $u$ and $u'$ in $g_2$, then $e_2$ is \emph{inserted} into $g_1$, denoted $\epsilon \rightarrow e_2$.
\end{enumerate}

If a vertex is deleted from (resp. inserted to) $g_1$ all its incident edges are automatically deleted (resp inserted). 

\subsection{Cost Function}
\label{subsec:Cost_Function}
 
The cost function assigns a cost to each edit operation incorporating domain-specific information about the similarity of objects. 
The substitution cost \( \beta \) is zero if the two vertices or edges have the same label. $c(u  \rightarrow v) = c(v \rightarrow u) = \beta$ /  $c(u \rightarrow \epsilon) = \theta$ / $c(\epsilon \rightarrow v) = \theta$
\section{Related Work}

\label{sec:related-works}

Graph-based representations address challenges such as image translation, rotation, and scaling~\cite{GED_survey_gao2010}, and GED is a widely used measure for exact and inexact graph matching. 
Since its introduction by Sanfeliu and Fu~\cite{sanfeliu1983distance}, GED has been solved using A-Star~\cite{bunke1983inexact} and improved heuristics such as the bipartite heuristic~\cite{riesen2007speeding, riesen2009approximate, Riesen2014_UB_LB}. Depth-First approaches~\cite{Zeina_abu_2015exact} reduce memory usage, and parallel B\&B implementations~\cite{dabah2019efficient} exploit multi-threading and load balancing to accelerate GED computation.

Exact methods remain impractical for large graphs due to the NP-hard nature of GED, motivating approximate strategies. These include parallel fixed-time Branch \& Bound~\cite{abu2018parallel, abu2016distributed}, Beam Search~\cite{neuhaus2006fast}, and our K-Best tree-based approach~\cite{kbest_DABAH2021}, these methods lack accuracy and are not scalable for large graphs.

FAST-GED addresses these limitations, by efficient algorithmic design exploiting efficiently GPU computing power. 

\section{FAST-GED: GPU-Accelerated Graph Edit Distance} 
\label{sec:k-best}
\begin{figure}[ht]
\hspace{0cm}
\centering
\begin{minipage}[t]{0.49\textwidth}
    \centering
    \includegraphics[width=1\linewidth]{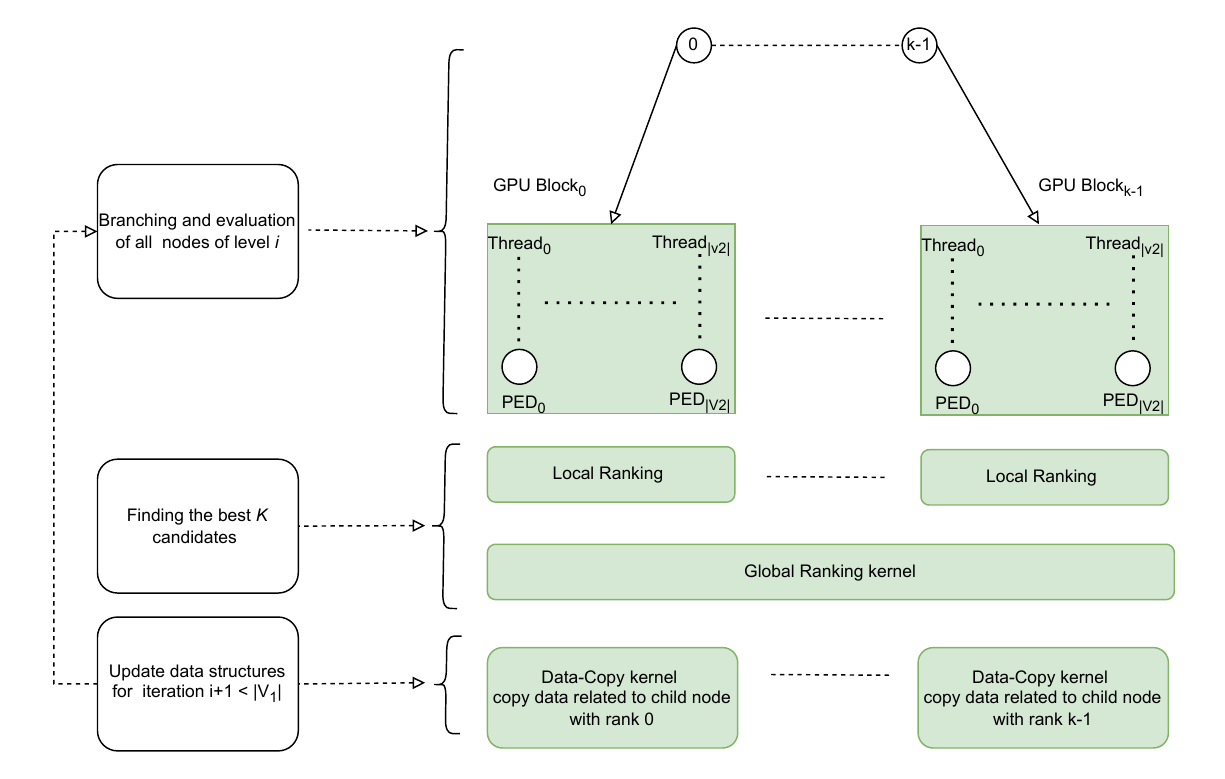}
    \caption{GPU-based parallelization of FAST-GED: Mapping of search tree expansion and ranking phases onto GPU blocks and threads.}
    \label{fig:fast-ged}
\end{minipage}
\hfill
\begin{minipage}[t]{0.48\textwidth}
\vspace{-4cm}
\begin{algorithm}[H]
\tiny
\caption{FAST-GED algorithm}
\label{alg:algorithm_BB_BF} 

\KwData{Attributed graphs $g_1 = (V_1, E_1, \alpha_1, \beta_1)$ and $g_2 = (V_2, E_2, \alpha_2, \beta_2)$ where $V_1= \{u_1,...,u_{|V_1|}\}$ and $V_2= \{v_1,...,v_{|V_2|}\}$}

\KwResult{Approximate graph edit distance and corresponding edit path}

\medskip
$List \leftarrow \{\text{initial\ problem}\}$; \\
$best\_path \leftarrow \text{null}$; 

\medskip
\ForEach{($v \in V_1$)}
{
  \ForEach{($node \in List$)}
  {
    \medskip 
    
    Generate $nd_j$, where $j \in [1,...,|R_{V2}|+1]$;
     
    \ForEach{successor($nd_j$) } 
    { 
       calculate $PED(nd_j)$;\\ 
       $list\_tmp \leftarrow \text{add\_node}(nd_j)$; 
    }  
  }
  $List \leftarrow \{\text{best } K \text{ nodes in } list\_tmp\}$
}
$best\_path \leftarrow \lambda $ (node with best $PED$ in List);\\
\Return \emph{best\_path};
\end{algorithm}

\end{minipage}
\end{figure}

The FAST-GED approach mimics the behavior of a Breadth-First Search (BFS) algorithm in traversing a search tree. The approach explores the tree level by level until reaching leaf nodes, performing both branching and evaluation for all nodes at each level before advancing to the next one. However, unlike traditional BFS, the FAST-GED approach selectively considers only the best $K$ nodes, based on evaluation criteria, for further exploration at the next level. Intuitively, the accuracy of this approach improves as the parameter $K$ increases. This selective exploration process involves two distinct phases: a branching phase, where all nodes at a given level undergo branching and evaluation, and a selection phase, where the best $K$ successor nodes are chosen for advancement to the next level based on their evaluation. These phases are repeated iteratively until the final level containing potential solutions is reached. Hence, the best solution is returned as an approximate solution for the GED problem. The best nodes selected at each level increase the likelihood of capturing the optimal path within the search space. \hyperref[alg:algorithm_BB_BF]{Algorithm~\ref*{alg:algorithm_BB_BF}} illustrates the different phases of this approach.

\subsection{GPU Architecture}
A Graphics Processing Unit (GPU) is an accelerator initially designed for gaming and optimized for parallel computation. GPU parallelism follows the Single Instruction Multiple Thread (SIMT) paradigm, where multiple threads execute concurrently. Threads are organized into blocks, and blocks form grids. Threads within a block can share data via shared memory and synchronize execution using barriers. GPUs also feature multiple memory hierarchies, including global, constant, texture, register and cache memory, each with specific access patterns. Making efficient use of these memory types is crucial for maximising GPU performance.


\subsection{Phase 1: Branching and Evaluation}
\subsubsection{Branching}
This phase generates successors of all nodes of a given level by decomposing each problem into several smaller sub-problems.  
The branching is based on the vertex-centric approach, where a vertex from the first graph $g_1$ is mapped to the remaining vertices from $g_2$ using substitution, deletion, and insertion operations.  
A search tree node $nd$ is characterized by:
\begin{itemize}
    \item A \textit{path} $\lambda = \{o_1, o_2, ..., o_k\}$ where $o_i$ represents past edit operations.
    \item Remaining vertices from $g_1$ and $g_2$ denoted by $R_{V1}$ and $R_{V2}$ respectively. 
\end{itemize}
In this way, the root node is defined as $|R_{V1}| = |V_1|$, $|R_{V2}| = |V_2|$, $\lambda = \emptyset$.
The branching on a search tree node $nd$ generates $|R_{V2}|$ new successors by performing the substitution of a vertex $v \in R_{V1}$ with all vertices in $R_{V2}$. In other words, each successor node $nd_j$ (where $j \in [1..|R_{V2}|]$) represents the mapping of a vertex $v \in R_{V1}$ to a vertex $u \in R_{V2}$ and is defined as follows: $ nd_j$ \{$R_{V1}(nd_j) = R_{V1}(nd)\setminus \{v\}$; $R_{V2}(nd_j) = R_{V2}(nd)\setminus \{u\}$; $ \lambda (nd_j) = \lambda (nd) \cup \{v \rightarrow u \}$\}. 
In addition, we add a new successor that represents the deletion of the vertex $v\in R_{V1}$.
$R_{V1}(nd_j) = R_{V1} (nd) \setminus \{v\}$; 
$R_{V2}(nd_j) = R_{V2} (nd) $;  
$ \lambda (nd_j) = \lambda (nd) \cup \{v \rightarrow \epsilon \}.$;

\subsubsection{Evaluation}
The evaluation process calculates the Partial Edit Distance (PED) for each successor node. i.e, Cost of all edit operations in $\lambda(nd_j)$: $PED (nd_j) =  \sum_{i=1}^{|y|} c(o_i) $.

\subsection{Phase 2: Selection} 
This step aims to select nodes more likely to contain promising paths. After evaluating all resulting nodes from the first phase, we select the $K$ best nodes based on their evaluations. Practically, this step involves ranking the nodes based on their evaluation and then removing all nodes ranked greater than $K$. This ranking process can be computationally intensive, particularly when dealing with a large number of nodes, which is challenging for real-time applications. To enhance the accuracy of this approach, a large value of $K$ is essential, as the differences in node evaluations are minimal, especially in the early stages of the search tree.   However, this will linearly increase the complexity. To overcome this issue, we leverage the computational power of GPUs.
\vspace{-0.4cm}
\subsection{GPU Implementation}
This section describes the GPU implementation of the FAST-GED algorithm. 
\vspace{-0.4cm}
\subsubsection{First Phase of FAST-GED}
To parallelize the first phase of the FAST-GED algorithm, which involves the generation and evaluation of successors on the GPU, we opted for the following mapping. As shown in \autoref{alg:FAST-GED-parallel}, each GPU block will handle the branching and evaluation of a node, such that each thread on the GPU will be responsible for generating and evaluating one successor. This involves the mapping of a vertex $v$ from the first graph $g_1$ to the remaining vertices in $g_2$ using the substitution operation. The total number of block threads is equal to the number of vertices in $g_2$; i.e., thread \textit{i} always performs the substitution operation using $u_i$. In addition, one thread will be responsible for the deletion operation. 
Note that vertex insertions are handled at the end of the search process: once all vertices of $g_1$ have been mapped or deleted, any remaining vertices in $g_2$ are inserted, and the corresponding costs are added.


\begin{algorithm}
\small
\caption{\small First Phase of FAST-GED on GPU} 
\label{alg:FAST-GED-parallel} 
\SetAlgoLined
\KwIn{Two graphs $g_1$ and $g_2$}
\KwOut{Successor nodes evaluated}
\textbf{Initialization:}\
Nb GPU blocks $\gets k$\;
Nb threads per block $\gets$ $|V_2|$ Nb vertices in $g_2$\;
\textbf{Parallel Execution:} \\
\ForEach{block $b$}{
    get vertex $v$ in $R_{V1}$ from $node_b$\;
    \ForEach{thread $t$ in block $b$}{
        \If{$u_t \in R_{V2}$}{
         substitute vertex $v$ with $u_t$ in $g_2$ ;
         $Imp\_cost = Implied\_Edge\_Cost()$ ;
         $PED = E(node_b) + c(v \gets u_t) + Imp\_cost$;
         }
    }
}
\end{algorithm}

The second part of this phase involves evaluating all successor nodes, which is done in two steps. The first step is common to all threads in a block, and the second step is specific to each thread. 
In the common first step, we calculate the implied edges following the mapping of vertex $v$ in $g_1$. This requires going through all previous edit operations $o_i$ in $\lambda$. For each edit operation $o_i$, we check if there is an edge between $v$ and the source or destination vertices of $o_i$. The results are stored in two vectors, $VFrom$ and $VTo$, in shared memory.

Next, in the thread-specific second step, each thread $t$ examines all edit operations in $\lambda$. For each $o_i$, we check if there is an edge between $u_t$ and the source or destination vertices of $o_i$. For each edit operation, we consider three cases: If an edge exists in $g_1$ ($VFrom[i] \ne \text{Null}$ or $VTo[i] \ne \text{Null}$) but not in $g_2$, we add an edge deletion cost. If an edge exists in $g_2$ but not in $g_1$, we add an edge insertion cost. If the edge exists in both graphs but with different weights, we add an edge substitution cost.

Finally, the partial edit distance is computed for each successor node evaluation, which is the sum of the parent partial edit distance, the cost of vertex mapping, and the implied edges cost. 

\subsubsection{Second Phase of FAST-GED}
The second phase of the FAST-GED algorithm involves selecting the best $K$ candidates for the next iteration. While a straightforward approach would be to perform sorting and select the top $K$ nodes, sorting algorithms can be computationally expensive and introduce significant overhead, particularly for scalability purposes. However, we do not require a sorted list; rather, we only need the top $K$ candidates in a non-sorted order. We do not perform a full sort of all candidates. Instead, we directly extract the top-K nodes using block-local selection and global thresholding. The final set of $K$ candidates is therefore unordered. 
Initially, we observed a lack of efficient algorithms capable of performing such operations on the GPU. Therefore, we propose the following approach to address this issue. 
After evaluation, we determine the local ranking of threads within each block using shared memory to reduce latency and facilitate data sharing among all threads in a block. The top $L$ threads from each block then update a global list of the best $L$ Partial Edit Distance (PED) values using atomic operations. Limiting the operation to the best threads helps mitigate the overhead associated with atomic operations.

Subsequently, a second kernel is employed to assign global rankings to threads. Threads with evaluations equal to the highest value are assigned the top ranks. Subsequent threads with evaluations equal to the second-highest value are ranked thereafter, continuing until the top $K$ candidates are identified.

Finally, we update the data structures with the values for the next iteration, thereby avoiding any host-device communication.
During experimentation, we set $L = 5$; however, we have the flexibility to adjust this value without a significant time increase. 


\subsection{Complexity Analysis}
In the following, we analyze the time and space complexity of FAST-GED. Let \( n = \max(|V_1|, |V_2|) \) and \( K \) be the number of nodes retained per level. Note that as $K \to \infty$, FAST-GED becomes equivalent to exhaustive breadth-first search and therefore converges to the optimal edit distance.

\textbf{Time complexity:} 
At level $i$ ($0 \leq i < n$), each of the $K$ retained nodes branches to  
$|RV_2| + 1$ ,\textit{i.e.},  $n+1$ successors in worst case (substitutions with all remaining vertices in $RV_2$, plus one deletion). 
Computing the partial edit distance for a successor requires iterating over the current 
edit path $\lambda$ (of size $\leq i \leq n$) to account for implied edges, yielding 
$\mathcal{O}(n)$ per successor. 
Sequentially this gives $\mathcal{O}(K \cdot n \cdot n) = \mathcal{O}(K n^2)$ per level and $\mathcal{O}(K n^3)$ overall across $n$ levels. 

When performing the parallelization on the GPU, we map one node to a GPU block where each thread handles one successor. This collapses the successor loop, and each block performs an $\mathcal{O}(n)$ shared pass over $\lambda$, so the per-node cost is $\mathcal{O}(n)$. 
With sufficient parallelism to process the $K$ nodes concurrently, the per-level time is 
$\mathcal{O}(n)$, leading to an overall complexity of $\mathcal{O}(n^2)$.
 
The top-$K$ selection has a linear complexity due to local rankings and global atomic operations rather than full sorting. It can become a bottleneck when $K$ is extremely large, in this case, approximate top-k selection may become an option. 

\textbf{Space complexity:} 
The algorithm space complexity is  $\mathcal{O}(K \cdot n)$. Indeed, each of the $K$ nodes at a given level stores an edit path $\lambda$ and two sets of remaining vertices, each of size $\mathcal{O}(n)$. Therefore, the total memory grows linearly with both $K$ and $n$. 
In practice, this trade-off between accuracy (through $K$) and memory cost is handled 
efficiently using GPU memory hierarchies, data reuse, and parallel reductions.


\section{Performance Evaluation}
\label{sec:Tests_and_experiments}
In this section, we evaluate the accuracy and scalability of {FAST-GED} using standard datasets and state-of-the-art methods. The source code and datasets are available in the FAST-GED repository\footnote{\url{https://gitlab.jsc.fz-juelich.de/dabah2/fast-ged/-/tree/main}}. We define $g_1$ as the source and $g_2$ as the target throughout all experiments for consistency. Vertex substitution, insertion, and deletion costs are set to 2, 4, and 4, respectively, while edge substitution, insertion, and deletion costs are 1, 2, and 2. These settings favor substitution over insertions and deletions to avoid trivial edit paths, but other cost functions are also supported. All experiments are performed on a JUWELS Booster node with four NVIDIA A100 GPUs and dual AMD EPYC 7702 CPUs (i.e. 48 cores). Unless stated otherwise, the parameter $K$ is fixed to 700{,}000.


We first validate FAST-GED by comparing its deviation from the optimal results of the \netx{} library~\cite{networkx}, a Python package for complex graph analysis that uses a DFS strategy with the bipartite heuristic~\cite{riesen2009approximate} to estimate future path costs. \hyperref[tab:approach_comparison]{Table~\ref*{tab:approach_comparison}} summarizes our results for small random graphs (ten vertices) across different graph densities, reporting average edit distance,  deviation from optimal \netx{} results, speedup, and the number of times FAST-GED reached optimal matches. 
Since \netx{} library requires several minutes per GED, 
 a 100 graph combinations for each density represents a good balance between computational effort and analysis. 
 
\begin{table}
\vspace{-0.6cm}
\centering
\small
\begin{tabular}{lc@{\,\ }c@{\,\ }c@{\,\ }c@{\,\ }c@{\,\ }c}
\toprule
\textbf{Method/Density} & \textbf{0.1} & \textbf{0.3} & \textbf{0.5} & \textbf{0.7} & \textbf{0.9} \\
\midrule
\netx{} & 17.95 & 19.21 & 19.97 & 20.83 & 22.51 \\
FAST-GED & 18.05 & 19.35 & 20.12 & 20.99 & 22.75 \\
\midrule
Deviation (\%) & 0.55 & 0.71 & 0.65 & 0.71 & 1.00 \\
Speedup & 26$\times$ & 38$\times$ & 48$\times$ & 50$\times$ & 55$\times$ \\
Optimal (\%) & 94/100 & 92/100 & 95/100 & 92/100 & 90/100 \\
\bottomrule
\end{tabular}
\caption{Comparison of \netx{} library optimal GED results and FAST-GED approach using random graphs at different densities.}
\label{tab:approach_comparison}
\end{table} 
 
\vspace{-1cm}
FAST-GED achieves less than 1\% deviation from \netx{}'s optimal results, while demonstrating substantial speedup improvements, ranging from $26\times$ to $55\times$. 
The speedup increase with graph density arises from the higher execution time of the \netx{} algorithm, whose bipartite heuristic has a complexity of $\mathcal{O}(n^3)$. In sparse graphs, the simpler bipartite structure allows near~$\mathcal{O}(n^2)$ performance and effective pruning, whereas in dense graphs, the structure becomes more complex, reducing pruning efficiency and increasing computational cost.
Unlike the \netx{}, FAST-GED's computational complexity remains agnostic to graph density, ensuring consistent performance across varying graph structures. Over 500 GED computations using random graphs, FAST-GED consistently achieves the optimal edit distance in more than \qty{90}{\percent} of cases, showcasing its near optimal performance.
 
\hyperref[tab:comparison]{Table~\ref*{tab:comparison}} presents the mean edit distance of FAST-GED and some of the best approximate approaches using medium-sized real-world datasets. 
The CMU dataset~\cite{cmu_faces}, created by Carnegie Mellon University's Robotics Institute. 
GREC and MUTA are subsets of the IAM graph database repository proposed in~\cite{dataset_riesen2008iam}.
The size column denotes graph size, and NB refers to the number of tested combinations.  We consider two baselines: the Beam Search (BS) method~\cite{neuhaus2006fast}, which limits its priority queue size to ten elements to balance accuracy and execution time; and DFS\_1~\cite{abu2016distributed}, one of the few approaches that scales. For each approach, we report the mean edit distance for all 55 graph combinations (lower is better). Optimal B\&B approaches cannot handle medium size graphs in a reasonable amount of time~\cite{dabah2022efficient}.
\begin{table}
\centering
\vspace{-0.6cm}
\small
\sisetup{detect-weight=true,detect-inline-weight=math}
\begin{tabular}{lS[table-format=2]S[table-format=2]S[table-format=2.1]S[table-format=3.1]S[table-format=3.1]S[table-format=3.1]S[table-format=3.1]S[table-format=3.1]}
\toprule
\textbf{Dataset} & \textbf{Size} & \textbf{NB} & {\makecell{\textbf{FAST-}\\{\bfseries GED}}} & \textbf{BS$_{10}$} & \textbf{DFS\_1} & {\makecell{\bfseries BS Time\\\bfseries(s)}} &{\makecell{\bfseries DFS time \\\bfseries(s)}} & {\makecell{\bfseries FAST-GED time\\\bfseries(s)}} \\ \midrule
\multirow{2}{*}{GREC} & 20 & 55 & \bfseries 32.0 & 37 & 39.8 & 01.7& 1.0&1.0 \\
 & \itshape Mix & 55 & \bfseries 25.4 & 36.9 & 28.3 & 01.4 & 1.0&1.0\\ \midrule
CMU & 30 & 55 & \bfseries 95.9 & 132.1 & 171.5 & 239.0 & 1.0&1.0\\ \midrule
\multirow{7}{*}{MUTA} & 20 & 55 & \bfseries 17.9 & 23.6 & 31.0 & 01.1 & 1.0&1.0\\
 & 30 & 55 & \bfseries 25.4 & 36.8 & 41.8 & 06.4 & 1.0&1.0\\
 & 40 & 55 & \bfseries 43.1 & 49.9 & 62.7 & 27.4 & 1.0&1.0\\
 & 50 & 55 & \bfseries 50.5 & 68.8 & 77.7 & 49.6 & 1.0&1.0 \\ 
 & 60 & 55 & \bfseries 65.4 & 79.5 & 86.6 & 250.0 & 1.0&1.0\\
 & 70 & 55 & \bfseries 73.6 &  & 113.5 &600 & 1.0&1.0 \\
 & \itshape Mix & 55 & \bfseries 86.6 & 140.8 & 100.9 & 297.0& 1.0&1.0\\
 \bottomrule
\end{tabular} 
\caption{Comparison of FAST-GED and state-of-the-art approximate methods on medium-sized real-world datasets.}
\label{tab:comparison}
\end{table}
\vspace{-1cm}

\autoref{tab:comparison} results confirm FAST-GED superiority, showing a lower mean edit distance compared to BS and DFS\_1 across all datasets. Regarding time to solution, DFS\_1 achieves the fastest execution time, followed by FAST-GED, with both approaches having an execution time of less than 1 second. Compared to BS$_{10}$, FAST-GED is up to 400 times faster for large datasets,  while maintaining higher accuracy. 
Indeed, retaining more candidate nodes per search-tree level increases the likelihood of FAST-GED capturing the optimal path. GPU parallelism enables this broader exploration at low cost, whereas CPU methods would require hours for similar results.

\pgfplotsset{
    every axis/.append style={
        width=6.2cm,
        height=5.2cm,
        tick label style={font=\footnotesize},
        label style={font=\footnotesize},
        legend style={font=\tiny},
        title style={font=\footnotesize},
    }
}

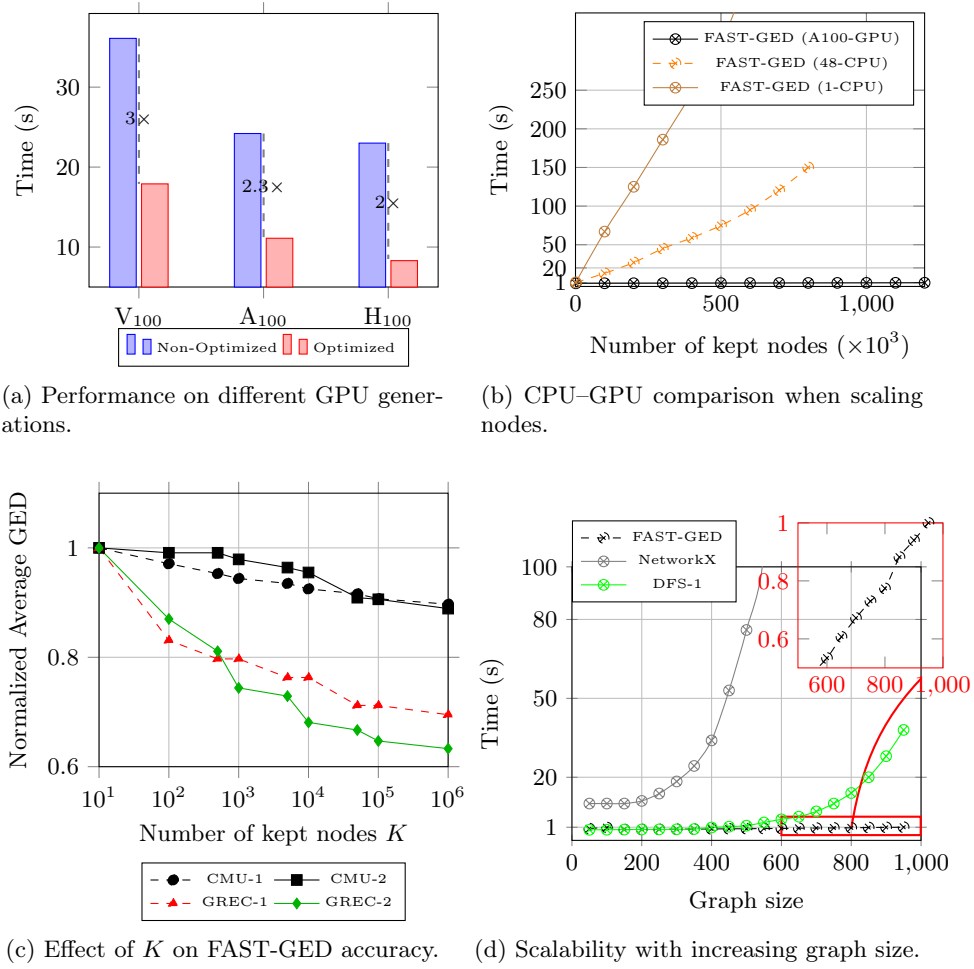
\begin{figure}[!ht]
    \centering

    \begin{subfigure}[t]{0.48\textwidth}
        \centering
        \begin{tikzpicture}
        \begin{axis}[
            ybar,
            symbolic x coords={V$_{100}$, A$_{100}$, H$_{100}$},
            xtick=data,
            ylabel={Time (s)},
            legend style={at={(0.5,-0.15)}, anchor=north, legend columns=-1},
            enlarge x limits=0.2,
            ymin=5,
            bar width=10pt
        ]
        \addplot coordinates {(V$_{100}$,36.1) (A$_{100}$,24.2) (H$_{100}$,23)};
        \addplot coordinates {(V$_{100}$,17.9) (A$_{100}$,11.1) (H$_{100}$,8.3)};
        \legend{Non-Optimized, Optimized}

        \draw[-, dashed, gray, thick] (axis cs:H$_{100}$, 23) -- (axis cs:H$_{100}$, 8.5);
        \node[color=black,font=\scriptsize] at (axis cs:H$_{100}$,15.5) {2$\times$};

        \draw[-, dashed, gray, thick] (axis cs:A$_{100}$, 24.2) -- (axis cs:A$_{100}$, 11.2);
        \node[color=black,font=\scriptsize] at (axis cs:A$_{100}$,17.5) {2.3$\times$};

        \draw[-, dashed, gray, thick] (axis cs:V$_{100}$, 36.1) -- (axis cs:V$_{100}$, 17.9);
        \node[color=black,font=\scriptsize] at (axis cs:V$_{100}$,26) {3$\times$};
        \end{axis}
        \end{tikzpicture}
        \caption{Performance on different GPU generations.}
        \label{fig:gpu}
    \end{subfigure}
    \hfill
    \begin{subfigure}[t]{0.48\textwidth}
        \centering
        \begin{tikzpicture}
        \begin{axis}[
            xlabel={Number of kept nodes ($\times 10^3$)},
            ylabel={Time (s)},
            grid,
            xmin=0, xmax=1200,
            ymin=-4, ymax=350,
            ytick={1,20,50,100,150,200,250},
            legend pos=north east,
            y label style={rotate=0},
            axis lines=box
        ]
        \addplot[black, mark=otimes, smooth, solid] coordinates {(1,0.035) (100,0.12) (200,0.175) (300,0.24) (400,0.29) (500,0.355) (600,0.41) (700,0.473) (800,0.54) (900,0.6) (1000,0.65) (1100,0.72) (1200,0.78)};
        \addlegendentry{FAST-GED (A100-GPU)};

        \addplot[orange, mark=otimes, smooth, dashed] coordinates {(1,1.215) (100,13) (200,27) (300,45) (400,59) (500,75) (600,95) (700,120.462) (800,150)};
        \addlegendentry{FAST-GED (48-CPU)};

        \addplot[brown, mark=otimes, smooth, solid] coordinates {(1,1.215) (100,67) (200,125) (300,186) (400,246) (500,309) (600,400) (700,438.462)};
        \addlegendentry{FAST-GED (1-CPU)};
        \end{axis}
        \end{tikzpicture}
        \caption{CPU–GPU comparison when scaling nodes.}
        \label{fig:cpu-gpu}
    \end{subfigure}

    \vspace{0.5cm}

    \begin{subfigure}[t]{0.48\textwidth}
        \centering
        \begin{tikzpicture}
        \begin{axis}[
            xlabel={Number of kept nodes $K$},
            ylabel={Normalized Average GED},
            grid=both,
            xmin=10, xmax=1e6,
            ymin=0.6, ymax=1.1,
            xmode=log,
            log basis x={10},
            xtick={10,100,1000,10000,100000,1000000},
            xticklabels={$10^1$,$10^2$,$10^3$,$10^4$,$10^5$,$10^6$},
            y label style={rotate=0},
            axis lines=box,
            tick align=outside,
            legend style={at={(0.5,-0.35)},anchor=north,legend columns=2}
        ]
        \addplot[black, mark=*, dashed] coordinates {(10,1.0) (100,0.971) (500,0.953) (1000,0.944) (5000,0.935) (10000,0.925) (50000,0.916) (100000,0.907) (1000000,0.897)};
        \addlegendentry{CMU-1};

        \addplot[black, mark=square*, solid] coordinates {(10,1.0) (100,0.991) (500,0.991) (1000,0.979) (5000,0.964) (10000,0.955) (50000,0.909) (100000,0.906) (1000000,0.889)};
        \addlegendentry{CMU-2};

        \addplot[red, mark=triangle*, dashed] coordinates {(10,1.0) (100,0.831) (500,0.797) (1000,0.797) (5000,0.763) (10000,0.763) (50000,0.712) (100000,0.712) (1000000,0.695)};
        \addlegendentry{GREC-1};

        \addplot[green!70!black, mark=diamond*, solid] coordinates {(10,1.0) (100,0.870) (500,0.811) (1000,0.744) (5000,0.729) (10000,0.681) (50000,0.667) (100000,0.647) (1000000,0.633)};
        \addlegendentry{GREC-2};
        \end{axis}
        \end{tikzpicture}
        \caption{Effect of $K$ on FAST-GED accuracy.}
        \label{fig:increas-k}
    \end{subfigure}
    \hfill
    \begin{subfigure}[t]{0.49\textwidth}
        \centering
     \begin{tikzpicture}
\begin{axis}[
    xlabel={Graph size},
    ylabel={Time (s)},
    grid,
    xmin=0,
    xmax=1000,
    ymin=-4,
    ymax=100,
    ytick={1,20,50,80,100},
    y label style={rotate=-00},
    axis lines=box,
    tick align=center ,
    legend style={at={(0.0,1.17)}, anchor=north west}
]
\addplot[black, mark=otimes, smooth, dashed] coordinates{  
    (50,0.45) (100,0.86) (150,0.127) (200,0.171)
    (250,0.200) (300,0.256) (350,0.288) (400,0.333)
    (450,0.390) (500,0.428) (550,0.482) (600,0.534)
    (650,0.607)  (700,0.672) (750,0.730) (800,0.774)
    (850,0.882) (900,0.940) (950,1.000)
};
\addlegendentry{FAST-GED};
\addplot[gray, mark=otimes, smooth, solid] coordinates { 
    (50,10) (100,10) (150,10) (200,11) (250,13.800)
    (300,18.400) (350,24.300) (400,34.000) (450,53.000)
    (500,76.000) (550,105.000) (600,200.000)
};
\addlegendentry{\netx{}};
\addplot[green, mark=otimes, smooth, solid] coordinates 
 { 
    (50,0.015) (100,0.054) (150,0.1) (200,0.141)
    (250,0.1700) (300,0.2) (350,0.3) (400,0.9)
    (450,1.2) (500,1.6) (550,2.8) (600,4) (650,5)
    (700,7) (750,10) (800,14) (850,20) (900,28)
    (950,38.000)
};
\addlegendentry{DFS-1};
\draw[red,thick] (axis cs:999,-2) rectangle (axis cs:600,5);


\draw[red,->,thick]
    ($(axis cs:800,4.5) - (0,0.1)$)
    to[bend left=20]
    (5.cm,2.5cm);
\end{axis}

\begin{axis}[
    at={(2.99cm,4.2cm)},  
    anchor=north west,
    width=3.5cm,
    height=3.5cm,
    xmin=500, xmax=1000,
    ymin=0.50, ymax=1,
    red
]

\addplot[
    black,
    mark=otimes,
    smooth,
    dashed
] coordinates{
 (500,0.428) (550,0.482) (600,0.534)(650,0.607)  (700,0.672) (750,0.730) (800,0.774)
    (850,0.882) (900,0.940) (950,1.000)
};

\end{axis}

\end{tikzpicture}
        \caption{Scalability with increasing graph size.}
        \label{fig:scal-size}
    \end{subfigure}

    \caption{Comprehensive performance evaluation of FAST-GED: 
    (a) performance across GPU generations, 
    (b) CPU–GPU scaling, 
    (c) effect of node parameter $K$ on accuracy, and 
    (d) scalability with graph size.}
    \label{fig:all}
\end{figure}

\autoref{fig:gpu} shows the performance of optimized and non-optimized FAST-GED across different GPU architectures. The optimized version is $2\times$ faster, with both versions using shared memory. 
The main optimization targets the final step of FAST-GED, which consists of preparing the data for the next iteration. Each thread previously copies parent-node data through three sequential loops, causing GPU-thread divergence and scattered memory accesses causing performance degradation. As a result, this step accounted for \qty{40}{\percent} of the total runtime. 
The optimized version introduces a new \textit{copy\_kernel}  in a block-wise fashion, where each thread has only three coalesced global memory accesses, reducing this step to only \qty{5}{\percent} of the total runtime. 

The A100 GPU achieves \qty{40}{\percent} better performance than V100, due to the improved architecture, memory bandwidth and core count. 
Similarly, the H100 (PCIe \qty{80}{\giga\byte}) yields a \qty{55}{\percent} performance gain compared to V100 and \qty{28}{\percent} over A100.  Indeed, the H100 has further architectural improvements, and FAST-GED did not fully benefit for it due to the synchronization and atomic operations needed to filter the top $k$ candidates at each level of the tree, suggesting future optimizations should target reducing this overhead.

\autoref{fig:cpu-gpu} illustrates the runtime of FAST-GED serial, multi-core CPU, and GPU-based (A100) versions as the $K$ increases using two small graphs with 20 vertices. The serial version runtime increases significantly due to the number of successors that need to be evaluated and sorting overhead. The multi-core CPU version generates and evaluates successors in parallel on 48 AMD CPU cores, resulting in a $4.5\times$ relative speedup. This version uses more memory, limiting kept nodes under 800,000. On the other side, mapping and optimizing FAST-GED fully onto the GPU and avoiding any host-device data transfer allows retaining million of nodes while keeping runtime under one second with up to $300\times$ faster runtime.

\hyperref[fig:increas-k]{Figure~\ref*{fig:increas-k}} illustrates the advantage of increasing the number of nodes retained at each level of the FAST-GED search tree. We use the 55 graph combinations for both the CMU and GREC (20) datasets in \autoref{tab:comparison} and report the average edit distance for all combinations. In addition, two cost-settings have been used for both the CMU and GREC datasets. 
 \textbf{Setting 1:} 
    $\{c_{\text{vsub}}=2,\, c_{\text{vdel}}=c_{\text{vins}}=4,\, c_{\text{esub}}=1,\, c_{\text{edel}}=c_{\text{eins}}=2\}$, 
   where substitutions are cheaper, favoring mapping-based edits. \textbf{Setting 2:} 
    $\{c_{\text{vsub}}=4,\, c_{\text{vdel}}=c_{\text{vins}}=12,\, c_{\text{esub}}=1,\, c_{\text{edel}}=c_{\text{eins}}=10\}$, where high insertion and deletion costs discourage structural changes. 

 Average edit distances are normalized by their values at $K=10$ for comparability. Across both datasets and cost settings, the same trend can be observed: increasing $K$  reduces normalized edit distance, demonstrating improved accuracy. 
Two phases are distinguish: a rapid improvement for $K=10$--$1000$, followed by slow decrease ($K>1000$) as FAST-GED converges to the optimal edit distance as $K \rightarrow \infty$.  
Indeed, $K$ serves as a tunable parameter that controls the trade-off between accuracy and efficiency, adjustable to the needs of each application.

\hyperref[fig:scal-size]{Figure~\ref*{fig:scal-size}} illustrates the scalability of FAST-GED~($K$=5000) as the graph size increases. Used graphs were randomly generated with an average density of 0.4 using \netx{}.  
FAST-GED  and DFS\_1 exhibit similar scalability performance for graphs up to 600 vertices. Beyond this point, FAST-GED maintains a nearly linear runtime increase exploiting lightweight GPU threads, following the $\mathcal{O}(n^2)$ complexity analysis predicted in Section 4.5. Whereas DFS\_1 shows a significant slowdown due to (i) the growing number of successors generated and evaluated sequentially and (ii) higher cost of traversing past edits in $\lambda$ to compute implied edge costs. 

Compared to the \netx{} approximate results, FAST-GED offers a faster and more scalable solution for GED computation, enabling its use in large-scale applications.

\section{GED Applications}
\label{sec:apps}
We demonstrate the effectiveness of {FAST-GED} through two relevant applications: Graph Classification and Neural Architecture Search (NAS).
\begin{figure}[ht]
\centering
\begin{subfigure}[t]{1\textwidth}
    \centering
    \resizebox{0.9\textwidth}{!}{
    \begin{tikzpicture}[
      node distance=0.5cm and 0.5cm,
      every node/.style={draw, minimum height=1cm, minimum width=0.5cm, align=center},
      arrow/.style={-Stealth},
      font=\small
    ]
    \node (search) {Search Space};
    \node[right=of search] (agent) {Agent};
    \node[right=of agent] (candidate) {Candidate\\Architecture};
    \node[right=of candidate] (eval) {Evaluation\\Strategy};
    \node[right=of eval] (optimal) {Optimal\\Architecture};
    \draw[arrow] (search) -- (agent);
    \draw[arrow] (agent) -- (candidate);
    \draw[arrow] (candidate) -- (eval);
    \draw[arrow] (eval) --  (optimal);
    \draw[arrow] (eval.south) -- ++(0,-1.5) -|  (agent.south);
    \end{tikzpicture}
    }
    \caption{Overview of Neural Architecture Search.}
    \label{fig:nas-overview}
\end{subfigure}

\begin{subfigure}[t]{0.49\textwidth}
    \centering
    \resizebox{\textwidth}{!}{
    \begin{tikzpicture}
    \begin{axis}[
        ybar,
        ymin=0, ymax=100,
        ylabel={Accuracy (\%)},
        symbolic x coords={GNN\_NDP, GNN\_MP, KNN\_GED},
        xtick=data,
        bar width=16pt,
        nodes near coords,
        nodes near coords align={vertical},
        enlarge x limits=0.3,
        grid=major,
        width=10cm,
        height=7cm,
    ]
    \addplot+[fill=blue!70] coordinates {(GNN\_NDP,78) (GNN\_MP,80) (KNN\_GED,75)};
    \end{axis}
    \end{tikzpicture}
    }
    \caption{Classification accuracy on the Mutagenicity dataset.}
    \label{fig:classification-accuracy}
\end{subfigure}
\hfill
\begin{subfigure}[t]{0.45\textwidth}
    \centering
    \resizebox{\textwidth}{!}{
    \begin{tikzpicture}
        \begin{axis}[
            title={Speedup in Log Scale},
            xlabel={NAS Graph Size},
            ylabel={Speedup Factor},
            xmin=9, xmax=13,
            ymin=1, ymax=1e5,
            xtick={9,10,11,12,13},
            ytick={1,10,100,1000,10000},
            ymode=log,
            log ticks with fixed point,
            legend pos=north west,
            grid=major,
            grid style=dashed,
        ]
        \addplot+[
            only marks,
            mark=*,
            error bars/.cd,
            y dir=both,
            y explicit,
            error bar style={red},
        ] coordinates {
            (9, 8.6)
            (10, 29.8)
            (11, 300)
            (12, 2140)
        };
        \addlegendentry{Measured Speedup}
        \end{axis}
    \end{tikzpicture}
    }
    \caption{FAST-GED speedup in architecture generation vs. \netx{}.}
    \label{fig:nas-speedup}
\end{subfigure}

\caption{Applications of FAST-GED. 
(a) Neural Architecture Search overview. 
(b) Graph classification accuracy using KNN\_GED. 
(c) Speedup in NAS graph generation compared to \netx{}.}
\label{fig:ged-applications}
\end{figure}
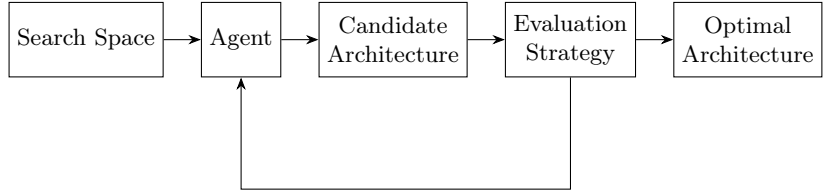
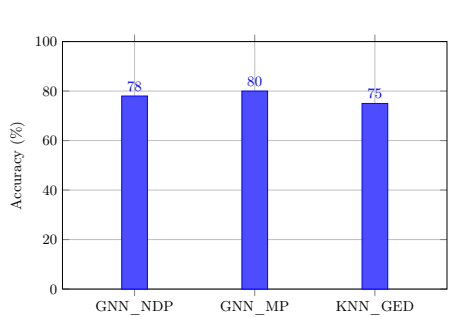
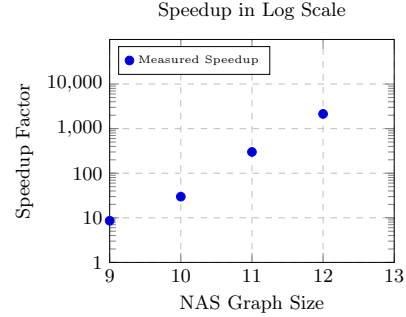

\subsection{Graph Classification}
Graph classification is a core machine learning task where models often struggle to capture structural similarity. 
Here, we combine FAST-GED with the K-Nearest Neighbor algorithm (\textbf{KNN\_GED}) to enable a simple yet effective classification method based on structural distances. 
We use the Mutagenicity dataset~\cite{dataset_riesen2008iam}, which classifies chemical compounds as mutagenic or non-mutagenic. 
Graphs are split into \qty{70}{\percent} training and \qty{30}{\percent} testing sets. 
Each test graph is assigned to the class of its nearest training graph in terms of GED. 

While computing thousands of pairwise GEDs using traditional approaches is prohibitive, taking weeks on large datasets, our GPU-accelerated framework reduces this to minutes, making KNN+GED practical and scalable. 
As shown in \autoref{fig:classification-accuracy}, KNN\_GED achieves accuracy comparable to the best advanced Graph Neural Networks (GNN\_MP~\cite{nouranizadeh2021maximum}, GNN\_NDP~\cite{bianchi2020hierarchical}) while remaining interpretable and easy to implement. 
Using uniform GED costs ($c_{\text{ins}}{=}c_{\text{del}}{=}2$, $c_{\text{sub}}{=}1$) and the nearest neighbor ($k{=}1$), the method attains \qty{75}{\percent} accuracy; tuning cost parameters can further improve results.

\subsection{Neural Architecture Search (NAS)}
Neural Architecture Search (NAS) is another machine learning field where GED can play an important role. As depicted in \autoref{fig:nas-overview}, the NAS goal is to explore the search space to automatically find the optimal architecture for a given problem that can outperform the manually designed architectures~\cite{zoph2016neural}.  Neural architectures are represented as directed graphs, and exploring the search space efficiently involves applying operations on the graphs, like crossover and mutation in evolutionary search. GED can be used in the evaluation phase to avoid evaluating similar tested architectures. This induces a huge gain since we avoid fitting a usually large amount of input data to the tested neural architecture, saving a lot of computing resources. 

The second interesting application is in generating new candidate architectures by an efficient crossover operation based on GED~\cite{qiu2023shortest}. Indeed, given two graphs,  the edit path that transforms one graph into another can be used to generate new architectures by applying the transformation path to a certain number of edit operations, resulting in a new graph that combines the best parts of both parents. In other words, a new architecture $C$ is generated from $A$ and $B$ by computing the GED edit path between them and applying half of its edit operations, producing a mixed graph of both. However, the final accuracy of the obtained model relies on both a near-optimal and fast GED computation, which was possible thanks to our open-source framework.

\autoref{fig:nas-speedup} shows the speedup obtained by FAST-GED compared to \netx{} when generating ten new architectures with varying graph sizes. 
Our framework achieves up to $10^3\times$ acceleration with less than 10\% deviation, well within the acceptable \qty{30}{\percent} margin~\cite{qiu2023shortest}. 
This drastic reduction in computation time—from hours to seconds—makes GED-based NAS strategies feasible at scale, enabling broader and more efficient exploration of neural architectures.

\section{Conclusion}
\label{sec:conclusion}
In this work, we presented FAST-GED, a high-performance framework for fast and scalable graph edit distance (GED) computation on GPUs. 
GED is a key metric in many applications, yet traditional methods struggle with large graphs due to their high computational complexity. 
FAST-GED addresses this challenge by combining tenable K-Best edit path search with an efficient GPU mapping that avoids any host-device data transfer, while efficiently leveraging GPU resources.  
Experimental results on both real-world and synthetic datasets demonstrate FAST-GED's efficiency, showcasing $26-55\times$ speedup over the \netx{} library for small graphs while consistently achieving better accuracy outperforming state-of-the-art methods on real-world datasets. 

Future directions focus on extending FAST-GED to process extremely large graphs with hundreds of thousands of vertices.

{
\bibliographystyle{splncs04}
\bibliography{ref}
}
\end{document}